\documentclass[conference]{IEEEtran}
\usepackage{amsmath}
\usepackage{booktabs}
\usepackage{graphicx}
\usepackage{url}
\usepackage{cite}
\usepackage{balance}
\usepackage[hidelinks]{hyperref}

\title{Federated KPI Aggregation with Local Privacy Constraints and Coordinated OLAP Roll-Ups}
\author{\IEEEauthorblockN{Felix Härer}\\\IEEEauthorblockA{
\textit{Center for Digital Trust} \\
\textit{University of Applied Sciences and Arts Northwestern Switzerland} \\
Dornacherstrasse 394, Basel, Switzerland \\
ORCID: 0000-0002-2768-2342 \\
E-Mail: felix.haerer@fhnw.ch
}
}

\begin{document}
\maketitle

\begin{abstract}
Cross-organizational KPI analysis can require aggregated results from independently administered data warehouses without centralizing detailed records. To limit disclosure from small local groups, a site can require a minimum number of contributing records before releasing a KPI value for a group. If each site applies this rule independently, different local contributions may be suppressed and the combined KPI can become incomplete. This paper investigates whether OLAP granularity can be coordinated before local query results are released so that all sites satisfy their local minimum-cohort rules and contribute at the same analytical level. The proposed method selects the least-generalized common roll-up accepted by every site. Site agents coordinate current capabilities and approved local execution across independently administered sites, with privacy checks and aggregate computation remaining deterministic local functions. In a four-site prototype, coordinated roll-up retained all contributions and reproduced the corresponding centralized result, compared with 78.3\% retained under independent suppression. A capability-change experiment showed fewer candidate checks and replies and no unsupported requests when current site-agent state was used. The results demonstrate the feasibility of federated KPI aggregation under local privacy constraints through coordinated granularity selection and agent-based coordination.

\end{abstract}


\section{Introduction}
Business intelligence derives key performance indicators (KPIs) from detailed operational data by aggregating measures over dimensions such as time, geography, customers, or products. Data warehouses and online analytical processing (OLAP) provide an established basis for these analyses: dimensional structures organize measures and attributes, and hierarchy levels support roll-up from detailed to coarser views, for example from city to region or from month to quarter \cite{chaudhuri1997,gray1997,kimball2013}.

When an analysis spans organizations, business units, or other independently administered data holders, the required data are distributed across separate systems. Privacy, confidentiality, or governance restrictions may require detailed records to remain under local control instead of being pooled in one warehouse. Federated analytics addresses this setting by executing computations where the data reside and exchanging derived results instead of raw records \cite{wang2024,srinivas2024}. Each independently administered data holder together with its local analytical warehouse is referred to as a \emph{site}.

Keeping records local does not by itself resolve disclosure risk. An OLAP query at a detailed hierarchy level can produce small groups at one site even when the corresponding group is sufficiently large across the federation. To limit disclosure from such small local groups, statistical disclosure control uses minimum cell-count rules and suppression \cite{hundepool2012}. A site can therefore require a minimum number of contributing records before releasing a KPI value for a group. When sites apply this rule independently, however, different local contributions may be suppressed. The resulting federation-wide SUM, mean, rate, or other KPI can then be incomplete even though every released local value satisfies the threshold.

Instead of suppressing local values independently, the proposed method uses the OLAP hierarchies to select, before local query results are released, a common roll-up at which all participating sites satisfy their local minimum-cohort rules with as little generalization as possible. Applying this method across independently administered sites also requires capabilities, authorization, and analytical tools to remain under local control. A coordinator agent and one site agent per participant therefore manage the coordination process. Site agents communicate current capabilities and authorization state, invoke approved deterministic functions, and accept or refuse requests as local conditions change. Privacy checks and numerical KPI computation remain the responsibility of deterministic functions.

These requirements lead to two research questions:

\smallskip
\noindent\textbf{RQ1.} How can a federation select the least-generalized common
OLAP roll-up that satisfies the local minimum-cohort threshold at every site?

\smallskip
\noindent\textbf{RQ2.} How can agents coordinate this selection across
independently administered sites and still keep capability, authorization, and
execution decisions local?
\smallskip

RQ1 is addressed by the roll-up selection method and its evaluation against independent suppression. RQ2 is addressed by an agent-based coordination architecture that keeps site-local state and approved tool invocation with the participating sites and allows unsupported requests to be refused and coordination to be replanned. Prototype and replication material are available online\footnote{\url{https://github.com/fhaer/Federated-Aggregation-Coordination-and-Privacy}}.

The remainder of the paper is organized as follows. Section~II discusses related work on federated analytics, OLAP privacy, multidimensional aggregation, and agent-based analytics. Section~III presents the coordinated aggregation method and architecture. Section~IV evaluates privacy--utility behavior and coordination under changing site capabilities. Section~V discusses the results, their implications, and further research.

\section{Related Work}
Federated analytics supports statistical and database-like computations over data that remain distributed among their owners \cite{wang2024,srinivas2024}. Recent work also addresses privacy protection within such systems. Cormode and Bharadwaj propose sample-and-threshold differential privacy for federated histograms \cite{cormode2022}, and Feldman \emph{et al.} study local pan-privacy for federated analytics \cite{feldman2025}. These approaches provide privacy guarantees beyond the minimum-cohort rule considered here and can complement the coordination of analytical granularity. 

The multidimensional data model and the OLAP roll-up operation are well established \cite{chaudhuri1997,gray1997}. Privacy-preserving OLAP has been investigated using perturbation and reconstruction over partitioned data \cite{agrawal2005}, inference control for published aggregates \cite{zhang2011}, and secure querying of encrypted warehouses \cite{fugkeaw2024}. A related line of work in statistical disclosure control uses cell thresholds and suppression to restrict sparse outputs \cite{hundepool2012}. Classical $k$-anonymity instead concerns indistinguishability of person-specific records in released microdata and commonly uses generalization and suppression \cite{sweeney2002}. In this paper, $k$ denotes a minimum cohort size for aggregate cells and serves as a local release criterion.

Multidimensional analysis also demonstrates the practical value of selecting analytical granularity. Camozzi \emph{et al.} use fact and dimension structures for blockchain analysis and explicitly exploit adaptable granularity in multidimensional queries \cite{camozzi2022}. This use of adaptable granularity supports treating hierarchy levels as an analytical design choice, which is central to the coordinated roll-up selection considered here.

Agent-based approaches have also been applied to federated analytics for query planning and workflow construction. LAFA uses a hierarchical multi-agent architecture to transform natural-language requests into executable federated analytics workflows \cite{ji2025}. The present work assigns agents a focused coordination role across independently administered sites. They maintain local operational state, invoke approved deterministic tools, and refuse unsupported or unauthorized requests. This division between distributed coordination and local execution is also consistent with prior architecture work on decentralized applications \cite{haerer2022exec,haerer2023pkia}.

\section{Coordinated Federated Aggregation}
\subsection{Sites, OLAP Levels, and Privacy Threshold}

Consider $N$ participating sites and an analytical query $Q$ that aggregates a numerical measure over one or more dimensions. For each grouped dimension $d$, the local warehouses provide compatible hierarchy levels ordered from detailed to coarse values, e.g. city $\rightarrow$ region $\rightarrow$ country. A roll-up $L$ is defined by selecting one hierarchy level for each grouped dimension.

The federation specifies a minimum cohort threshold $k$, and each site may enforce a stricter local threshold $k_i$. Its effective threshold is therefore $\kappa_i=\max(k,k_i)$. The federation threshold is transmitted with each candidate, allowing the site to evaluate $\kappa_i$ locally without transmitting $k_i$. Let $c_i(g,L)$ denote the number of local privacy units contributing to a non-empty output group $g$ at site $i$ under roll-up $L$. A candidate is accepted when
\begin{equation}
\mathrm{safe}_i(L) \iff
\min_{g:c_i(g,L)>0} c_i(g,L) \geq \kappa_i.
\label{eq:safety}
\end{equation}

For a site with no non-empty output groups under $L$, $\mathrm{safe}_i(L)$ is defined as true and the site returns an empty local result. In the current prototype, the privacy unit is a fact row. If a person can contribute multiple fact rows, the threshold can instead be evaluated using distinct subject identifiers.

Before evaluating privacy, the coordinator obtains the hierarchy levels and measures that the sites currently support and are authorized to release. Their combined capabilities define the executable candidate set $C(Q)$. Within this set, candidates are ordered by an information-loss value $I(L)$ that measures the normalized hierarchy distance from the requested levels. The selected roll-up is
\begin{equation}
L^*=\arg\min_{L\in C(Q)} I(L)
\quad\text{s.t.}\quad
\mathrm{safe}_i(L)\;\;\forall i=1,\ldots,N.
\label{eq:selection}
\end{equation}
Because candidates are evaluated in increasing order of $I(L)$, the first level accepted by every site is the least-generalized executable candidate under the chosen loss measure.

\subsection{Roll-Up Coordination Example}
Table~\ref{tab:example} illustrates the selection for three sites with a federation threshold $k=10$. The requested KPI is \emph{SUM(revenue) by city and exact age}, using the geography hierarchy $G_0$=city, $G_1$=region, $G_2$=country and the age hierarchy $A_0$=exact age, $A_1$=decade, $A_2$=age band. Each site entry gives the minimum cohort size among its non-empty output groups, so a candidate is safe only if all three values are at least 10.

\begin{table}[t]
\centering
\caption{Example with three sites and threshold $k=10$. Values are local minimum cohort sizes.}
\label{tab:example}
\footnotesize
\setlength{\tabcolsep}{3.0pt}
\begin{tabular}{llrrrc}
\toprule
Candidate & OLAP grouping & Site 1 & Site 2 & Site 3 & Result \\
\midrule
$L_0$ & city $\times$ exact age   & 4  & 12 & 6  & unsafe \\
$L_1$ & region $\times$ exact age & 8  & 15 & 9  & unsafe \\
$L_2$ & city $\times$ decade      & 7  & 13 & 8  & unsafe \\
$L_3$ & region $\times$ decade    & 14 & 18 & 12 & select \\
\bottomrule
\end{tabular}

\vspace{3pt}

\footnotesize Geography: city $\rightarrow$ region $\rightarrow$ country.\\
Age: exact $\rightarrow$ decade $\rightarrow$ age band.
\end{table}

\subsection{Agent-Based Coordination}
\begin{figure*}[t]
\centering
\includegraphics[width=1\textwidth]{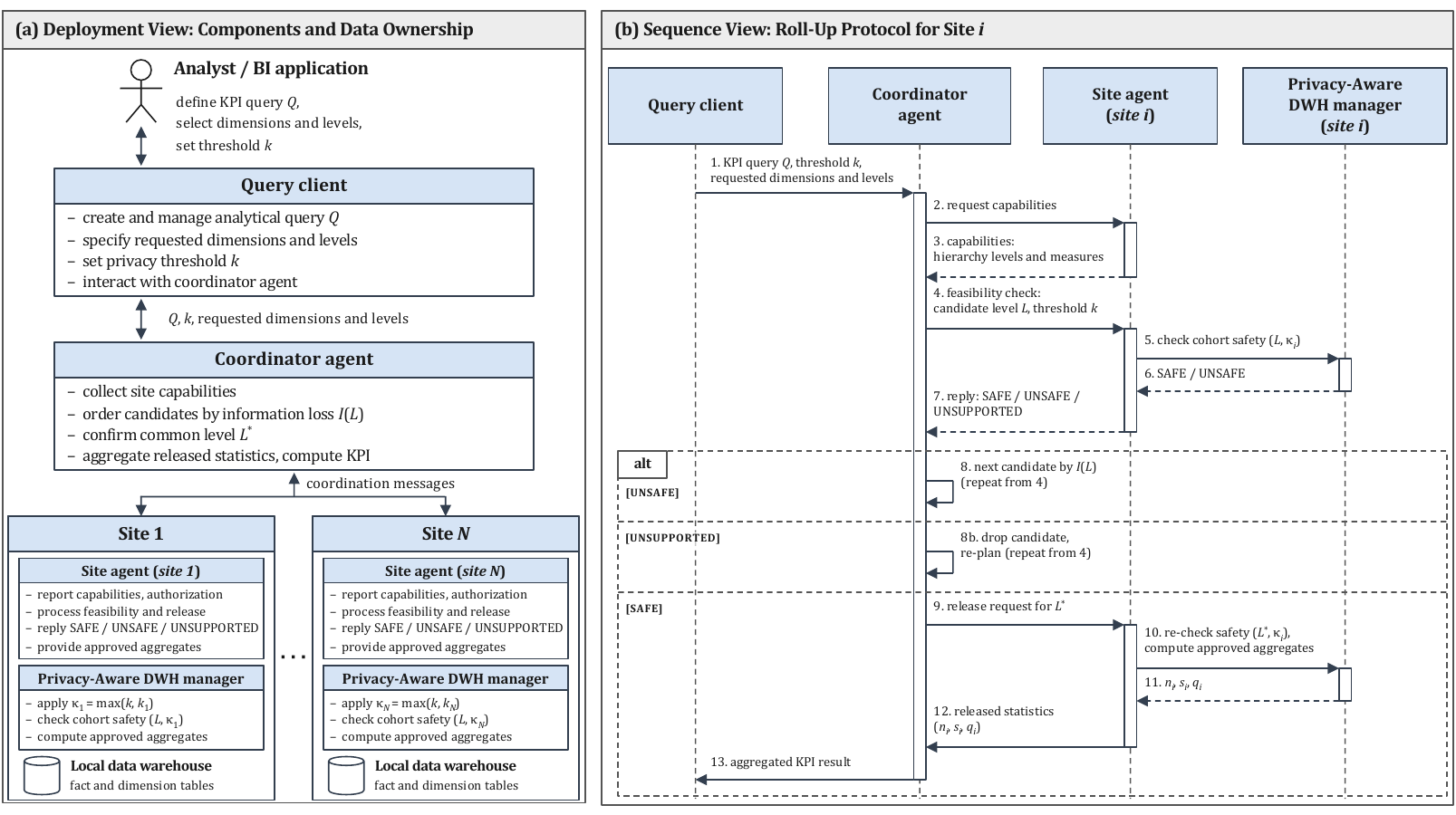}
\caption{Coordinated federated KPI aggregation. (a)~Deployment view: query client and coordinator agent, and $N$ independently administered sites, each retaining its own site agent, privacy-aware DWH manager, and local data warehouse. (b)~Sequence view for one site~$i$: capability exchange, evaluation of a candidate roll-up $L$ against the effective threshold $\kappa_i=\max(k,k_i)$, and a final deterministic threshold check after the common level $L^*$ is confirmed and before release of the additive sufficient statistics $n_i=\mathrm{COUNT}$, $s_i=\mathrm{SUM}(x)$, and $q_i=\mathrm{SUM}(x^2)$. Site agents reply \texttt{SAFE}, \texttt{UNSAFE}, or \texttt{UNSUPPORTED}. No record-level data are transferred.}
\label{fig:arch}
\end{figure*}

The coordinator agent proposes candidates in increasing $I(L)$. For $L_0$, Sites 1 and 3 contain at least one local group below $k$. For $L_1$, Sites 1 and 3 still fail. For $L_2$, the same sites fail again. Candidate $L_3$ is the first level for which all three sites satisfy Eq.~\eqref{eq:safety} and is therefore selected. The cohort sizes in Table~\ref{tab:example} serve only to illustrate this decision. During coordination, sites return \texttt{SAFE}, \texttt{UNSAFE}, or an operational refusal such as \texttt{UNSUPPORTED}.

Once a candidate has been selected, every site repeats the deterministic threshold check immediately before releasing its local sufficient statistics. It then returns additive sufficient statistics for each selected group. In the prototype, these are $n_i=\mathrm{COUNT}$, $s_i=\mathrm{SUM}(x)$, and $q_i=\mathrm{SUM}(x^2)$. From these statistics, the coordinator reconstructs the global count and sum by addition, the mean as $\sum_i s_i/\sum_i n_i$, and the population variance from the corresponding sums of squares. No record-level data are transferred.

Figure~\ref{fig:arch} combines deployment and sequence views of the distributed architecture. In the deployment view, the query client submits query $Q$, threshold $k$, dimensions, and requested hierarchy levels to the coordinator agent. Based on capabilities reported by the sites, the coordinator orders candidate roll-ups by increasing information loss $I(L)$ and requests the release of local aggregate statistics once a common level has been established. Each independently administered site retains its site agent, privacy-aware DWH manager, and local data warehouse. The sequence view follows this interaction for one site, including capability exchange, feasibility replies (\texttt{SAFE}, \texttt{UNSAFE}, or \texttt{UNSUPPORTED}), and release of the approved statistics $n_i$, $s_i$, and $q_i$ after selection of $L^*$. Capability and authorization state remain local to the site, and deterministic functions perform the cohort check and aggregate computation.

This division of responsibilities assigns agents to coordination and approved tool invocation. They maintain workflow state, exchange current capability information, and invoke approved local tools. Requests and query specifications pass deterministic validation before execution. Cohort checks, the guarded release of local aggregate statistics, and KPI arithmetic remain deterministic local functions.

\subsection{Prototype Implementation}
The research prototype is implemented in Python with one independent SQLite star schema per site. Each schema contains a \texttt{fact\_sales} table and customer, product, and date dimensions, with hierarchies for geography, age, product, and time. Site-specific geographic concentrations, product preferences, and purchase frequencies produce different local sparsity patterns and therefore different cohort sizes for the same analytical query.

The coordination layer is independent of a specific agent runtime or message transport. The repository includes an implementation adapter that uses the Hermes agent framework to realize the coordinator and site-agent roles and the Buzz agent communication platform to carry structured coordination messages \cite{hermes2026,buzz2026}. The adapter covers capability exchange, roll-up proposals and replies, aggregate requests, and the release of local sufficient statistics. The deterministic privacy and aggregation functions are shared by the experiment harness and the implementation adapter. This realization provides the agent coordination of RQ2, with privacy decisions and KPI computation remaining in the deterministic local functions.

\section{Evaluation}
\subsection{Experimental Setup}
The privacy--utility experiment uses four sites with synthetic data, 450 customers and 5,000 fact rows per site, three random seeds, six two- or three-dimensional query shapes, and thresholds $k\in\{3,5,10,20\}$. Site-specific distributions create different local sparsity patterns, so that the same grouping can be sufficiently large at one site and fall below the cohort threshold at another.

As a baseline, \emph{local suppression} evaluates the originally requested grouping and removes every local cell below $k$ before combining the remaining values. Coordinated roll-up instead selects a common level that satisfies the threshold at all sites and combines their contributions at that level. The comparison uses retained-row coverage, relative SUM error, roll-up loss, and the number of candidate checks. Suppression error is measured at the requested level, and coordinated roll-up is compared with the centralized result at the selected level. This reflects the trade-off between the two approaches: suppression preserves the requested granularity by omitting contributions, and coordinated roll-up preserves the contributions by adjusting granularity.

\subsection{Privacy--Utility Results}
\begin{figure}[t]
\centering
\includegraphics[width=0.94\columnwidth]{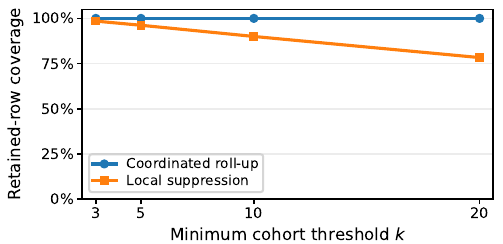}
\vspace{-0.2em}
\caption{Mean retained-row coverage: as $k$ increases, local suppression omits more contributions, while coordinated roll-up retains all at a common level.}
\label{fig:coverage}
\end{figure}

Table~\ref{tab:results} summarizes the results over 18 query cases per threshold. Under independent suppression, the retained-row contribution decreases as $k$ increases, as shown in Figure~\ref{fig:coverage}, resulting in less complete combined KPI. At $k=20$, 78.3\% of fact-row contributions remain and the mean relative SUM error is 21.6\%. Coordinated roll-up retains all contributions and reproduces the corresponding centralized result at the selected level. Achieving this requires coarser analytical levels as the threshold increases, with mean normalized roll-up loss rising from 0.346 at $k=3$ to 0.418 at $k=20$.

\begin{table}[t]
\centering
\caption{Privacy--utility results.}
\label{tab:results}
\footnotesize
\begin{tabular}{lrrrr}
\toprule
Method & $k$ & Coverage & SUM err. & Roll-up loss \\
\midrule
Suppression & 3  & .985 & .015 & .000 \\
Suppression & 5  & .962 & .038 & .000 \\
Suppression & 10 & .900 & .099 & .000 \\
Suppression & 20 & .783 & .216 & .000 \\
Coordinated roll-up & 3  & 1.000 & .000 & .346 \\
Coordinated roll-up & 5  & 1.000 & .000 & .384 \\
Coordinated roll-up & 10 & 1.000 & .000 & .395 \\
Coordinated roll-up & 20 & 1.000 & .000 & .418 \\
\bottomrule
\end{tabular}
\par{\raggedright\footnotesize \vspace{1.7mm} Means over 18 query cases per $k$. Note: the SUM error for roll-ups is taken relative to the centralized-equivalent answer at the selected level.\par}
\end{table}

The hierarchy search also remained limited in the evaluated workload, with an average of 5.6 candidates tested for $k=3$ and 7.0 for $k=20$. These values depend on the implemented dimensions and query shapes and do not constitute a general complexity bound. For the tested structures, they show that the common level can be identified with a small number of local feasibility checks.

\subsection{Coordination under Capability Changes}
A second experiment examines the coordination role of the site agents when local capabilities change. Four sites are assigned different supported hierarchy levels and local vocabulary, and the coordinator initially plans from a stored capability snapshot. After this snapshot is created, one site's available geography level changes from detailed geography to country level. The site agent then refuses proposals that are no longer executable under its current capability state.

With the stale snapshot, the coordinator tests five candidates, receives 20 replies, and encounters three \texttt{UNSUPPORTED} refusals (see Fig.~\ref{fig:arch}(b)) before finding an executable level. Using current capability information from the site agents reduces this to two candidates and eight replies, with no unsupported requests. Table~\ref{tab:coordination} summarizes the comparison. The experiment therefore shows the value of keeping current capability state with the participating sites and incorporating it into coordination when local conditions change.

\begin{table}[t]
\centering
\caption{Coordination after a site capability change.}
\label{tab:coordination}
\footnotesize
\begin{tabular}{lrrr}
\toprule
Capability information & Candidates & Replies & Unsupported \\
\midrule
Stored snapshot & 5 & 20 & 3 \\
Current site-agent state & 2 & 8 & 0 \\
\bottomrule
\end{tabular}
\end{table}

The coordination experiment uses a deterministic harness to isolate the effect of capability changes and make the state transitions reproducible. The prototype also contains an optional implementation adapter for the Hermes agent framework and the Buzz agent communication platform, together with the corresponding structured messages \cite{hermes2026,buzz2026}. This setup tests how current site-agent state affects coordination without introducing network or LLM latency into the measurements. The measurements concern the coordination logic and its response to changing capability state.

\section{Discussion and Outlook}

The experiments show a trade-off for federated business intelligence. Independent suppression preserves the requested granularity, but different sites may suppress different cells and thereby omit contributions from the combined KPI. Coordinated roll-up instead selects the least-generalized level that satisfies the release rule at every site. At this level, all contributions are retained and the result is identical to the corresponding centralized result. The trade-off is therefore expressed as a controlled change in analytical granularity instead of incomplete aggregation.

The minimum-cohort rule provides an explicit, locally enforceable release criterion for KPI groups with few contributing records \cite{hundepool2012}. Because it can be evaluated deterministically at each candidate OLAP level, the criterion can be incorporated directly into the selection of analytical granularity across independently administered sites. It is defined over aggregate cells and is distinct from record-level $k$-anonymity \cite{sweeney2002}. Additional privacy mechanisms, such as differential privacy and controls against inference from repeated queries, can be combined with the coordination method where stronger protection is required.

The interpretation of the cohort threshold also depends on the chosen privacy unit. In the current prototype, the threshold counts contributing fact rows, which may include multiple rows from one person. To count persons rather than fact rows, the local safety check can instead use distinct subject identifiers. This changes the cohort calculation without changing the coordination protocol or roll-up selection method.

Agent coordination addresses the operational independence of the participating sites. Each site agent maintains current capability and authorization state, invokes approved local functions, and can refuse requests that are no longer executable. In the capability-change experiment, using current site-agent state reduced the search from five candidates and 20 replies to two candidates and eight replies and eliminated unsupported requests. Keeping current operational state with the participating sites therefore improves coordination when local capabilities change. Privacy checks and aggregate computation remain deterministic throughout the process. 

Further work may extend the evaluation to subject-level privacy units, larger public or organizational datasets, live deployments, and repeated-query controls.

Overall, the results demonstrate how federated KPI aggregation can satisfy local privacy constraints based on minimum cohort sizes by coordinating OLAP granularity before local query results are released, avoiding incomplete results otherwise caused by independent suppression. Agent-based coordination keeps operational state with the participating sites and supports coordination when local capabilities change. Privacy enforcement and aggregate computation remain with separate deterministic local functions.

\section*{Acknowledgment}
OpenAI ChatGPT was used to assist in editing and proofreading. All content was reviewed and approved by the author, who takes full responsibility for the final manuscript.

\vspace{0.89em}
\balance
\bibliographystyle{IEEEtran}
\bibliography{fedkolap_references}

\end{document}